\documentclass[useAMS,usenatbib]{mn2e}

\usepackage[totalwidth=480pt,totalheight=680pt]{geometry}
\usepackage{aas_macros, epsfig}
\newcommand{\SFR}{{M$_{\sun}\,$yr$^{-1}$}}

\newcommand{\ltsim}{\mbox{{\raisebox{-0.4ex}{$\stackrel{<}{{\scriptstyle\sim}}\,$}}}}
\newcommand{\um}{~$\mu$m}

   \title[H-ATLAS: quasar hosts with Herschel]{\emph{Herschel}-ATLAS: The link between accretion luminosity and star formation in quasar host galaxies\thanks{\emph{Herschel} is an ESA space observatory with science instruments provided by European-led Principal Investigator consortia and with important participation from NASA}}
   \author[Bonfield et al.]{D. G. Bonfield$^{1}$\thanks{Email (DGB): d.bonfield@herts.ac.uk},  
M. J. Jarvis$^{1}$, %yes
M. J. Hardcastle$^{1}$, %yes 
A.~Cooray$^{2}$, %yes
E.~Hatziminaoglou$^{3}$, %yes
\newauthor
R.~J.~Ivison$^{4,5}$, %yes
M.~J.~Page$^{6}$, %yes
J.~A.~Stevens$^{1}$, %yes
G.~de~Zotti$^{7,8}$, %yes
R.~Auld$^{9}$, %yes %%%%%%%%%%%%%% after submission %%%%%%%%%%%%
M.~Baes$^{10}$, %yes
\newauthor
S.~Buttiglione$^{7}$, %yes
A.~Cava$^{11}$, %yes 
A.~Dariush$^{9,12}$, %yes
J.~S.~Dunlop$^{5}$, %yes
L. Dunne$^{13}$, %yes
S. Dye$^{9}$, %yes
\newauthor
S. Eales$^{9}$, %yes
J.~Fritz$^{10}$, %yes
R.~Hopwood$^{14}$, %yes
E.~Ibar$^{4}$, %yes
S.~J. Maddox$^{13}$, %yes
M.~J.~{Micha{\l}owski}$^{5}$, %yes
\newauthor
E.~Pascale$^{9}$, %yes
M.~Pohlen$^{9}$, %yes
E.~E.~Rigby$^{13}$, %yes
G.~Rodighiero$^{7}$, %yes
S.~Serjeant$^{14}$, %yes
\newauthor
D.~J.~B.~Smith$^{13}$, %yes
P.~Temi$^{15}$, %yes
P.~van~der~Werf$^{5,16}$ %yes
\\~\\
$^{1}$Centre for Astrophysics Research, Science \& Technology Research Institute, 
University of Hertfordshire, Hatfield, AL10 9AB, UK \\
$^{2}$Dept. of Physics \& Astronomy, University of California, Irvine, CA 92697, USA\\
$^{3}$ESO, Karl-Schwarzschild-Str. 2, 85748 Garching bei M{\"{u}}nchen, Germany\\
$^{4}$UK Astronomy Technology Centre, Royal Observatory, Edinburgh,
EH9 3HJ, UK\\
$^{5}$Scottish Universities Physics Alliance, Institute for Astronomy, University of Edinburgh, Royal Observatory, Edinburgh, EH9 3HJ, UK\\
$^{6}$Mullard Space Science Laboratory, University College London, Holmbury St. Mary, Dorking, Surrey RH5 6NT, UK\\
$^{7}$INAF -- Osservatorio Astronomico di Padova, Vicolo
 Osservatorio 5, I-35122, Padova, Italy\\
$^{8}$SISSA, Via Bonomea 265, I-34136 Trieste, Italy\\
$^{9}$School of Physics \&\ Astronomy, Cardiff University, Queen's
 Buildings, The Parade, Cardiff, CF24 3AA, UK \\
$^{10}$Sterrenkundig Observatorium, Universiteit Gent, Krijgslaan 281
 S9, B-9000 Gent, Belgium\\
$^{11}$Instituto de Astrof\'isica de Canarias (IAC) and Departamento de
 Astrof\'isica de La Laguna (ULL), La Laguna, Tenerife, Spain\\
$^{12}$School of Astronomy, Institute for Research in Fundamental Sciences
(IPM), PO Box 19395-5746, Tehran, Iran\\
$^{13}$Centre for Astronomy and Particle Theory, The School of
 Physics \&\ Astronomy, Nottingham University, University Park
 Campus, \\Nottingham, NG7 1HR, UK \\
$^{14}$Department of Physics \&{} Astronomy, The Open University, Walton Hall, Milton Keynes, MK7 6AA, UK\\
$^{15}$Astrophysics Branch, NASA/Ames Research Center, MS 245-6,  Moffett Field, CA 94035, USA\\
$^{16}$Leiden Observatory, Leiden University, P.O. Box 9513, NL-2300 RA Leiden, The Netherlands\\
}

\begin{document}
\date{Received Month dd, yyyy; accepted Month dd, yyyy}
\pagerange{\pageref{firstpage}--\pageref{lastpage}} \pubyear{2010}
\maketitle
\label{firstpage}

\begin{abstract}
We use the science demonstration field data of the {\em Herschel}-ATLAS to study how star formation, traced by the far-infrared {\em Herschel} data, is related to both the accretion luminosity and redshift of quasars selected from the Sloan Digital Sky Survey and the 2SLAQ survey. By developing a maximum likelihood estimator to investigate the presence of correlations between the far-infrared and optical luminosities we find evidence that the star-formation in quasar hosts is correlated with both redshift and quasar accretion luminosity.  Assuming a relationship of the form $L_{\rm IR} \propto L_{\rm QSO}^{\theta}(1+z)^{\zeta}$, we find $\theta = 0.22 \pm\, 0.08$ and $\zeta = 1.6 \pm\, 0.4$, although there is substantial additional uncertainty in $\zeta$ of order $\pm 1$, due to uncertainties in the host galaxy dust temperature.  We find evidence for a large intrinsic dispersion in the redshift dependence, but no evidence for intrinsic dispersion in the correlation between $L_{\rm QSO}$ and $L_{\rm IR}$, suggesting that the latter may be due to a direct physical connection between star formation and black hole accretion.  { This is consistent with the idea that} both the quasar activity and star formation are dependent on the same reservoir of cold gas, so that they are both affected by the influx of cold gas during mergers or heating of gas via feedback processes.

\end{abstract}


%
%________________________________________________________________

\section{Introduction}

AGN activity is now widely
believed to be an important phase in the evolution of every massive galaxy in
the Universe. This belief stems firstly from the discovery of a
relatively tight correlation between the mass of a galaxy's bulge and the mass
of its central supermassive black hole 
\citep[SMBH; e.g.][]{magorrian1998, fm2000, gebhardt2000},
which implies that build-up of the stellar mass and the SMBH mass are causally
connected \citep[although see also][who show that averaging of parameters by mergers 
can also produce such a correlation]{jahnke2010}.

From a more theoretical perspective, AGN have become a key
ingredient for semi-analytic models of galaxy formation 
\citep[e.g.][]{benson2003, cole2000, guideroni1998, granato2001}. 
The over-production of stars at
the bright end of the galaxy luminosity function by these models has led to the
inclusion of an extra source of heating or ``AGN feedback'' 
\citep[see e.g.][]{bower2006, croton2006, granato2004}. 
However, it is still
unclear what kind of AGN-driven feedback is the most important, e.g. two
{ complementary} mechanisms are described by \citet{croton2006}: ``quasar-mode''
feedback, { proposed to be effective during episodes of} 
efficient accretion of cold gas, and ``radio-mode''
feedback, { caused by the relatively inefficient accretion of hot gas 
(in the absence of cold gas) when
an AGN would have lower optical luminosity, but would also be
generating radio jets.}  Thus, it is now clear that if we are to understand
galaxy formation and evolution, we must also obtain a clear picture of
AGN, their environments and the co-evolution of SMBHs and
their host galaxies.

{
It is difficult to study the host galaxies of distant quasars, due to the nuclear quasar emission overwhelming the stellar emission at optical and near-infrared wavelengths.} However, {\em Hubble Space Telescope} and high spatial-resolution  ground-based imaging have enabled the investigation of the stellar populations of quasar hosts 
\citep[e.g.][]{Bahcall1994, McLure1999, Dunlop2003, Kotilainen2009},
demonstrating that most luminous quasars reside in massive elliptical galaxies.
{ A number of studies have attempted to determine the star-formation properties of quasar host galaxies, with some using optical colours \citep[e.g.][]{Sanchez2004} or spectroscopy \citep[e.g.][]{Nolan2001,silverman2009,trichas2010}.  Some information on the host galaxies of AGN has been gleaned from powerful radio galaxies, where the quasar nucleus is obscured due to the putative dusty torus \citep[e.g.][]{antonucci1993}. Again these studies show that powerful AGN reside in massive galaxies 
\citep[e.g.][]{Jarvis2001, McLure2004, Seymour2007} and there appears to be a trend where the most powerful exhibit signs of recent star formation \citep[e.g.][]{BC2008, Herbert2010}. However, optical colours and emission lines can be contaminated by the quasar, so that its contribution must be modelled to estimate the star-formation rate, and all of these studies suffer from the uncertainty arising from having very little information about how much star formation is obscured by dust, and thus how much of the emission due to young stars is reprocessed to far-infrared and sub-millimetre wavelengths.

Spectroscopy with the {\em Spitzer Space Telescope} has enabled the determination of star-formation rates in individual AGN via mid-infrared emission lines, which are much less susceptible to dust, \citep[e.g.][]{schweitzer2006,netzer2007,ms2008, lutz2008,Shi2009, trichas2009,lutz2010}. However, the relatively low sensitivity of {\it Spitzer}'s infrared spectrograph (IRS) has meant that only small, severely flux-limited samples of objects could be investigated, making the precise relationship between star-formation rate and quasar luminosity difficult to robustly determine.
}

(Sub)millimetre instruments such as SCUBA and MAMBO have provided important information on
{ reprocessed} emission from { star formation} in distant quasars and radio galaxies,
e.g. photometric surveys revealing luminous ($\sim10^{13}$ L$_{\sun}$)
far-infrared emission { from a number of radio galaxies and
quasar hosts, making it clear that major (SFR $\sim1000$\SFR) episodes of
star formation can coexist with powerful AGN }
\citep[e.g.][]{archibald2001,  priddey2003hiz, omont2003, page2004, ms2009}.
Furthermore, sub-mm imaging has revealed
overdensities of sources in the vicinity of individual high-z AGN,
plausibly belonging to the same large-scale structures 
\citep{stevens2003, priddey2007,stevens2010MNRAS}. { Conversely, 
cross-correlating SCUBA
data with deep X-ray surveys, \citet{alexander2005} also show a large
fraction of sub-mm galaxies contain buried AGN, while \citet{laird2010} find that
the X-ray AGN fraction in sub-mm galaxies is 20--29 \%.  
}

The main drawback of the sub-mm studies to date is that, due to the high background
associated with ground-based observing and the small number of pixels in 
existing instruments, it has not been possible to obtain
very large samples of objects. 
A recent study by \citet[hereafter SH09]{SH09} 
\defcitealias{SH09}{SH09} % cite by alias using \citetalias, \citepalias
attempted to overcome this shortfall by combining data from a wealth of \emph{Spitzer} and \emph{ISO} observations at mid- and far-infrared wavelengths to investigate how obscured star formation in the hosts of quasars evolved over cosmic time and also how this may be related to the bolometric output of the AGN. They found evidence for an increase in far-infrared luminosity as a function of both cosmic epoch and quasar luminosity.  However, they measured the correlation between far-infrared and quasar luminosity using wide redshift bins, which can potentially confuse the effect of redshift with that of quasar luminosity, since the latter evolves strongly with redshift in flux-density-limited samples.  In addition, their work extrapolated the far-infrared flux from mid-infrared wavelengths using SED templates, and it would clearly be preferable to measure the far-infrared flux directly.

The recently-launched ESA {\emph{Herschel}} space observatory \citep{pilbratt} 
offers an unprecedented combination of sensitive detectors, large (3.5-metre diameter) collecting area, and low background, which makes it possible, for the first time, to conduct deep surveys of large areas of the sky at far-infrared/sub-mm wavelengths.

In this paper we present a study of the far-infrared fluxes of known
quasars in science demonstration observations of the {\emph{Herschel}} Astrophysical 
Terahertz Large Area Survey \citep[H-ATLAS;][]{eales2010}.  This is the first single sample
of quasars measured in the sub-mm which is large and diverse enough to directly determine a
relationship between far-infrared (cool dust) luminosity and optical (quasar) luminosity, 
without using infrared information at wavelengths short enough to be possibly contaminated by warm 
dust in the AGN torus.  While \citet{serjeant2010} determine a relationship between optical and far-infrared luminosity using this data in combination with infrared luminosities derived from shorter-wavelength \emph{Spitzer} and \emph{ISO} data, here we use only the H-ATLAS data.  

Section \ref{sec:data} describes the H-ATLAS observations used.  Section \ref{sec:method} explains how we construct our quasar sample, determine far-infrared fluxes for them, and find a maximum-likelihood fit to the relationship between far-infrared luminosity, optical luminosity, and redshift.  The results of this fitting are discussed in Section \ref{sec:results} and conclusions are presented in Section\ref{sec:conc}.  

\section{Observations} \label{sec:data}
%\subsection{The {\emph{Herschel}}-ATLAS}
When completed, H-ATLAS will provide photometry for
 550 square degrees of the sky in five 
far-infrared/sub-mm bands. The 5-$\sigma$ depths at wavelengths of 
100, 160, 250,
350, and 500\um{} are 132, 126, 32, 36 and 
45~mJy respectively.  

%H-ATLAS is based on parallel scan mode observations with {\emph{Herschel}}.
%This provides data simultaneously from both the PACS (Poglitsch et
%al. \cite{pacs}) and SPIRE (Griffin et al. \cite{spire}) instruments.
%The time-line data reduced using {\tt HIPE}. Maps from the SPIRE data
%were produced using a naive mapping technique after removing
%instrumental temperature variations from the time-line data (Pascale
%et al \cite{spiremaps}). Noise maps were generated by using the two
%cross-scan measurements to estimate the noise per detector pass, and
%then for each pixel the noise is scaled by sqrt(number of detector
%passes). Maps from the PACS data were produced using photproject
%(Ibar et al \cite{pacsmaps}).

In this paper, we use only data from \emph{Herschel}'s SPIRE instrument \citep{spire}, 
at 250, 350, and 500\um, since the PACS 100 and 160\um{} data \citep{pacs}
are not deep enough to provide useful constraints on the  infrared luminosities of the
 quasars we are studying \citep{pacsmaps}.  The longer SPIRE wavelengths are also less likely to be 
contaminated by direct emission from the warm dust of the AGN torus.  
This means that the emission
can be more reliably attributed to cool dust heated by star-formation processes, and 
its spectral energy distribution (SED) approximated by a simple greybody.  

This paper is also restricted in area to the H-ATLAS science demonstration (SD) field, which
covers approximately 16 square degrees -- around 3 per cent of the total survey area. 
This SD data has been publicly released, and both the maps and catalogues can be retrieved  from the H-ATLAS webpage\footnote{http://www.h-atlas.org}.  

\section{Method}\label{sec:method}
%\subsection{Quasar sample}
Our quasar sample consists of all objects in the field which are identified as quasars
in the seventh data release of the SDSS Quasar Catalog \citep{qsodr7}, or the 2dF-SDSS 
LRG (luminous red galaxy) and Quasar spectroscopic catalogue \citep[2SLAQ;][]{2slaq}.  After removing 
the known blazar HB0906+015, since its far-infrared emission will not be 
dominated by star formation \citep[see e.g.][]{gonzalez-nuevo2010}, 
we find 372 quasars in the  whole SV field, with a reasonably
conservative cut on the area (by eye) to avoid using the noisier data at the edges 
of the field.  

%Explain that we have BH masses from Mg2 and C4 lines from SDSS and 2SLAQ spectra.  
%Obviously also have redshifts; can use optical magnitudes in SDSS to get approximate Lqso.  
%Eddington ratio is function(Lqso)/function(Mbh).

Of the 372 quasars, only 29 have matches to objects in the 5-$\sigma$ H-ATLAS catalogue \citep{cats, crossids}.  To improve our statistics, we measure fluxes directly from the SPIRE maps 
described by \citet{spiremaps} at the  position of each
 quasar.  To do this, we take a point measurement 
of the flux in the nearest pixel to the SDSS position on each
 PSF-filtered and background-subtracted map.  

\begin{figure}
\includegraphics[angle=0,width=0.48\textwidth,clip]{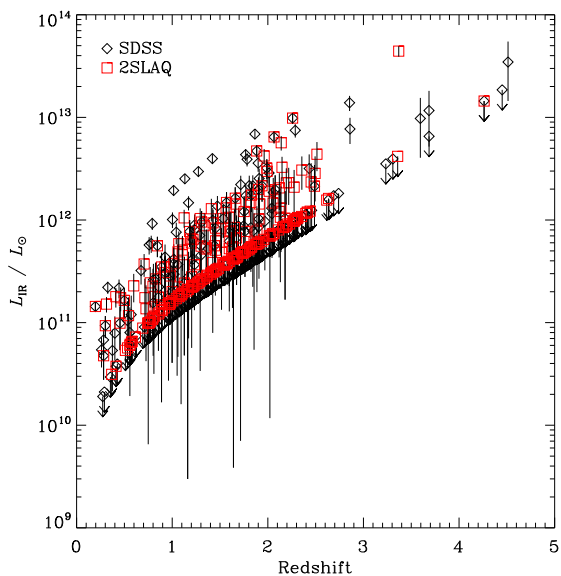}
\caption{Estimated total infrared luminosities (calculated by integrating the best-fitting greybody with $T=30$~K and $\beta=1.5$ over the 8--1000\um{} wavelength range) for our sample of spectroscopically confirmed quasars, plotted as a function of redshift.  Objects with spectra from SDSS are shown as black diamonds; those with 2SLAQ spectra have red squares.  Error bars are plotted where $L_{\mathrm{IR}}$ is greater than the uncertainty in the estimate, otherwise 1-$\sigma$ upper limits are shown using arrows. }\label{fig:lirz}
\end{figure}

To account for source confusion, the 
most likely flux from the object (which is used to produce Figure \ref{fig:lirz}) is obtained by subtracting the mean pixel value in the map.  However, when we perform a maximum-likelihood fit for the coefficients of Equation \ref{eq:lz}, as described below, there is an implicit assumption that the measurement noise is symmetric about zero.  The confusion noise is very significantly skewed, and can cause false correlations to be found in random data, so for the fluxes used in these fits we symmetrise the noise distribution by subtracting the flux in a different randomly selected pixel from each flux measurement.  This increases the overall noise level, so we average the likelihoods from a number of random background realisations in order to reduce the noise level in the final result.

%, and subtract the mean pixel value in the map
%from this flux measurement to account for source confusion.  
%The  noise is measured in the same way from 
%similarly-filtered noise maps, which were generated by \citeauthor{spiremaps} using the two
%cross-scan measurements to estimate the noise per detector pass, and
%scaled by $\surd$(number of detector passes) for each pixel.  

We also construct a comparison sample of fluxes drawn from random positions in the maps, with their backgrounds subtracted in precisely the same way as the real fluxes, and are then used to produce maximum-likelihood estimates in the same way as the real data.
Comparing these estimates with those from the real data allows us to increase our confidence that any trend detected in the data is due to the influence of the 
quasars, and not merely a result of the noise properties of the maps, which have a significant impact due to our inclusion of individual noisy objects instead of using binned data.

To compare the AGN luminosity of the quasars with that from star formation, we 
make the assumption that the optical light is dominated by the AGN, while the 
far-infrared/sub-mm is dominated by the cool dust heated by star formation.  
This latter assumption 
is consistent with the results of \citet{hatz10}, who fit models with 
AGN and star-formation contributions to the mid- and far-infrared emission from AGN
hosts and find that star formation dominates the far-infrared fluxes.  Their result 
depends on the assumption that the dusty torus heated directly by an AGN has a 
fairly simple, approximately toroidal geometry, as described by \citet{fritz2006}. 
The infrared SEDs of AGN 
can, however, be fully reproduced by more complex clumpy torus models  
\citep[e.g.][]{nenkova2008}, so, at least until these models are tested by 
high-spatial-resolution imaging of AGN tori,
we cannot completely rule out the possibility of 
a contribution to far-infrared fluxes from dust heated directly by the AGN.
However, the fact that \citet{hatz10} find that the SPIRE colours of AGN are 
indistinguishable from those of normal galaxies suggests that the AGN does not 
make a strong direct contribution at these wavelengths.

\begin{figure}
\includegraphics[angle=0,width=0.48\textwidth,clip]{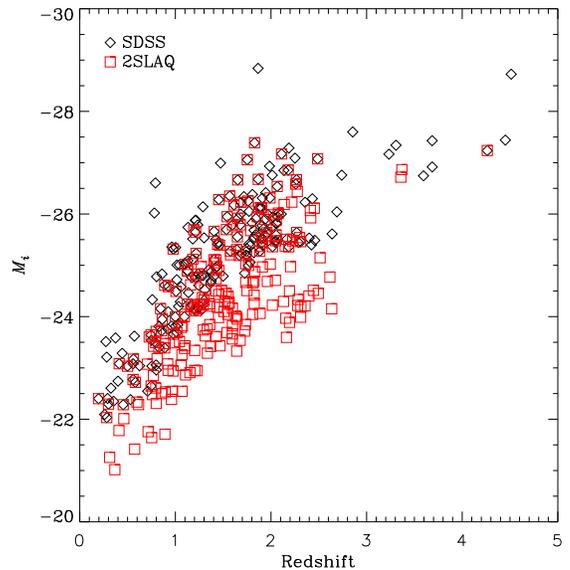}
\caption{Absolute $i$-band magnitudes (calculated as described in the text) for our sample of spectroscopically confirmed quasars, plotted as a function of redshift.  Objects with spectra from SDSS are shown as black diamonds; those with 2SLAQ spectra have red squares. }\label{fig:lumz}
\end{figure}

All of the quasars have spectroscopic redshifts from their respective surveys, 
and optical photometry in the $ugriz$ bands from the SDSS \citep{sdss2000}.  We use these 
data to estimate the quasar luminosity, following \citeauthor{qsodr7} and others in using
the absolute $i$-band magnitude, $M_i$, as a proxy for the luminosity of the central 
engine.  This is calculated from the 
observed $i$ magnitude (after the subtraction of Galactic extinction)
and the redshift, by assuming a power-law SED ($S_{\nu}\propto \nu^{\alpha}$) with spectral index 
$\alpha=-0.5$ and assuming a $\Lambda$CDM cosmology with $\Omega_m=0.3$, 
$\Omega_\Lambda=0.7$, and $H_0=70\,\mathrm{km\,s^{-1}\,Mpc^{-1}}$.  
Figure \ref{fig:lumz}
 shows the distribution of optical luminosities of our sample
as a function of redshift, indicating the source of the spectroscopic 
redshift (SDSS or 2SLAQ, or both).  From this we can see that the 2SLAQ
measurements in this field allow the inclusion of quasars at least one 
magnitude fainter (at $z<3$) than the SDSS spectroscopic survey probes.  This
greater range in optical luminosity is crucial for decoupling optical 
luminosity and redshift, and gives our study an advantage over similar work 
by \citet{hatz10}.  { We note, however, that the range of bolometric luminosities
spanned by our quasar sample is $10^{45}$ erg s$^{-1} < L_{\mathrm{bol}} < 10^{48}$ 
erg s$^{-1}$, so all of the objects are firmly in the quasar (as opposed to Seyfert)
regime. Thus, if there are two ``modes'' of accretion
with different feedback properties,
this study will only be relevant to understanding the ``quasar mode'', 
making it usefully complementary to the work of \citet{shao2010}, which uses different
{\it Herschel} data to probe the star-formation rates in lower-luminosity 
X-ray-selected AGN.}

As mentioned earlier, the SPIRE photometry at 250, 350, 
and 500\um{} is unlikely to be contaminated by the AGN directly for even 
the higher redshift quasars in the sample, and is deeper than the H-ATLAS
PACS photometry \citep{pacsmaps}, so we use only the SPIRE
fluxes to constrain the overall infrared luminosity.  To make the results 
directly comparable with each other, we fit a simple isothermal grey dust model, with 
flux density 
\begin{equation}
S_\nu \propto \nu^{(3+\beta)} / (e^{h\nu/kT)} - 1)  
\end{equation}
where $h$ and $k$ are Planck's and Boltzmann's constants, $\nu$ is 
rest-frame frequency, and where we assume fixed values for both 
temperature $T=30\,$K and $\beta=1.5$.  
We then integrate the greybody curve
between rest-frame wavelengths of 8 and 1000\um{} to obtain an estimate of 
$L_\mathrm{IR}$.  (This wavelength range is chosen for convenience and for 
comparison with other studies, and is not intended to imply that the cold dust 
represented by the greybody dominates the SED at wavelengths as short as 8\um{}.)  
We estimate the uncertainty on 
$L_\mathrm{IR}$ by increasing it until the $\chi^2$ value for the fit to the data increases by 
unity.

Since we include far-infrared flux measurements at our quasar positions regardless of whether there is a significant detection in the far-infrared, fitting the normalisation
of a fixed template is the only robust measurement possible.  This procedure does not account for the possibility of dust temperature 
or emissivity evolution either with redshift or quasar luminosity, which may well
 be important effects to consider.  However, we are able to use the well-detected objects in our sample to address the suitability of our choice of $T$ and $\beta$.  

To check our choice of greybody temperature, we fit greybody SEDs to the SPIRE fluxes of individual quasars with good detections (signal-to-noise $> 3$) in all three SPIRE bands, with $\beta$ fixed at 1.5 and $T$ allowed to vary.  The best-fitting temperatures for eight out of the nine well-detected quasars were very similar, with mean 32.4 K and a standard deviation of 2.8 K, while one outlier fits a temperature of $58.5_{-7.5}^{+14.0}\,$K.  The eight quasars span a redshift range of $1.1 < z < 2.3$ and a range in infrared luminosity (estimated by integrating the greybody between 8 and 1000 $\mu$m) of $6\times10^{12}<L_{\rm{IR}}/L_{\sun} < 2\times 10^{13}$.  They are consistent with the temperatures of galaxies as a whole as determined by \citet{amblard2010}, who find a mean temperature of $28\pm8$~K for galaxies detected at a significance of 3-$\sigma$ in the H-ATLAS SDP field, and a mean temperature of $30\pm9$~K for a compilation of galaxies in H-ATLAS and a number of other far-infrared surveys.  This consistency is unsurprising, since \citet{elbaz2010} also found (using deeper data from \emph{Herschel}) that the dust temperatures of radio-quiet AGN hosts were not significantly different from the normal galaxy population.  Thus we consider that a temperature $T=30$~K is likely to be a reasonable choice for our quasar sample, but is not completely certain, and we return to this point later.  

The outlying object has substantial uncertainties on its temperature, so we do not consider it an indication of a serious problem with our use of a single temperature for all objects. However, the fact that it is at a higher redshift ($z= 3.4$) and slightly higher luminosity ($L_{\mathrm{IR}}/L_{\sun} =  3\times10^{13}$) than the other objects means that it is broadly consistent with the evolution of dust temperature, to higher temperatures at higher redshifts and luminosities, seen in several previous studies \citep{chapman2005,kovacs2006,michalowski2010,amblard2010} and others.

We do not attempt to use a luminosity- or redshift-dependent dust temperature in this work.  Increasing the dust temperature serves to shift the greybody peak to shorter wavelengths, which would result in a higher luminosity fit to our data (which are mostly longward of the peak).  Thus, if the temperature of the dust in quasar hosts does increase with redshift and/or luminosity, then we will systematically underestimate $L_{\mathrm{IR}}$ for objects at high redshift and $L_{\mathrm{IR}}$.  However, since the trend in mean temperature found by \citeauthor{amblard2010} is weak compared with the dispersion around the relation, we do not expect this to be a very significant effect.  

To get an indication of the effect of temperature (and its uncertainty) on the parameters we derive, we perform fits to all objects using both $T=30$~K and $T=45$~K, where we have selected 45~K because it has a similar peak wavelength to the average $z>4$ quasar SED constructed by \citet{priddey2001} (who find parameters of $T = 41 \pm 5$ with $\beta = 1.95 \pm 0.3$).  We also note that, while $\beta$ may not in fact be 1.5, the luminosity estimates obtained using $\beta=2$ (using the best-fitting temperature at this $\beta$) differ by less than 1 per cent, so we consider that the selection of a value of $\beta$ is broadly degenerate with the choice of $T$, at least for the purpose of estimating infrared luminosity from SPIRE fluxes.  

Following \citetalias{SH09}, we assume that the dependence of the 
infrared luminosity on the optical quasar luminosity and redshift 
can be expressed as 
power laws, i.e.:
\begin{equation}  \label{eq:lz}
L_{\mathrm{IR,model}} = A\, L_{\mathrm{QSO}}^{\theta} \, (1+z)^{\zeta}
\end{equation}
with intrinsic (Gaussian)
dispersion in $\theta$ and $\zeta$ of $\Delta{\theta}$ and 
$\Delta{\zeta}$ respectively.  

Unlike \citeauthor{SH09}, who stack infrared data to look for trends in 
bins of redshift and optical luminosity, we fit for the 
power law coefficients of
equation (\ref{eq:lz}), and the normalisation $A$, by finding the peak in the 
likelihood distribution calculated over all data simultaneously.  
Our approach has the substantial disadvantage that variations in the power laws as a function of 
redshift or luminosity will be hidden from us.  However, under the assumption that the values of $\zeta$ and $\theta$ are the same at all redshifts and luminosities, our method enables us to make the most sensitive possible determination of these parameters, since we do not lose information through binning.  Additionally, we are immune to the potentially confusing effects of correlations within bins.     

We incorporate the dispersions $\Delta{\theta}$ and 
$\Delta{\zeta}$ by combining them with the uncertainty $\sigma_i$ 
on each datapoint.  
The likelihood, Prob(data$|\theta,\zeta,\Delta{\theta},\Delta{\zeta},A$) 
at a given point in parameter space is then given by 
\begin{equation}
\mathrm{Prob(data}|\theta,\zeta,\Delta{\theta},\Delta{\zeta},A)= 
\prod_i 
  \frac{e^{-\chi_i^2/2}}
       {\sqrt{2\pi(
           \sigma_i^2 + \sigma_{\theta,i}^2 + \sigma_{\zeta,i}^2)
       }}    
\end{equation}
where
\begin{equation}
\chi_i^2 = \frac{(L_{\mathrm{IR,data},i} - L_\mathrm{IR,model,i})^2}
                {\sigma_i^2 + \sigma_{\theta,i}^2 + \sigma_{\zeta,i}^2}
\;\mathrm{,}
\end{equation} \begin{equation}
\sigma_{\theta,i} =  
    L_\mathrm{IR,model,i} \ln L_{\mathrm{QSO},i}\,\Delta{\theta} 
\;\mathrm{,\,and}
\end{equation} \begin{equation}
\sigma_{\zeta,i} =
    L_\mathrm{IR,model,i} \ln(1+z)\,\Delta{\zeta}
\;\mathrm{.}\end{equation}

The prior space on the parameters $\theta$, $\zeta$, $\Delta{\theta}$, 
$\Delta{\zeta}$, and $A$, over which the likelihood was evaluated, was 
chosen iteratively to enclose more than 99.999 per cent of the likelihood.  The
priors were linear in $\theta$, $\zeta$, $\Delta{\theta}$, and 
$\Delta{\zeta}$, and logarithmic in $A$, with the 
limits $\Delta{\theta} \geqslant 0$ and $\Delta{\zeta}\geqslant 0$.  

\section{Results and Discussion}\label{sec:results}

\begin{figure*}
\includegraphics[angle=0,width=0.95\textwidth,clip]{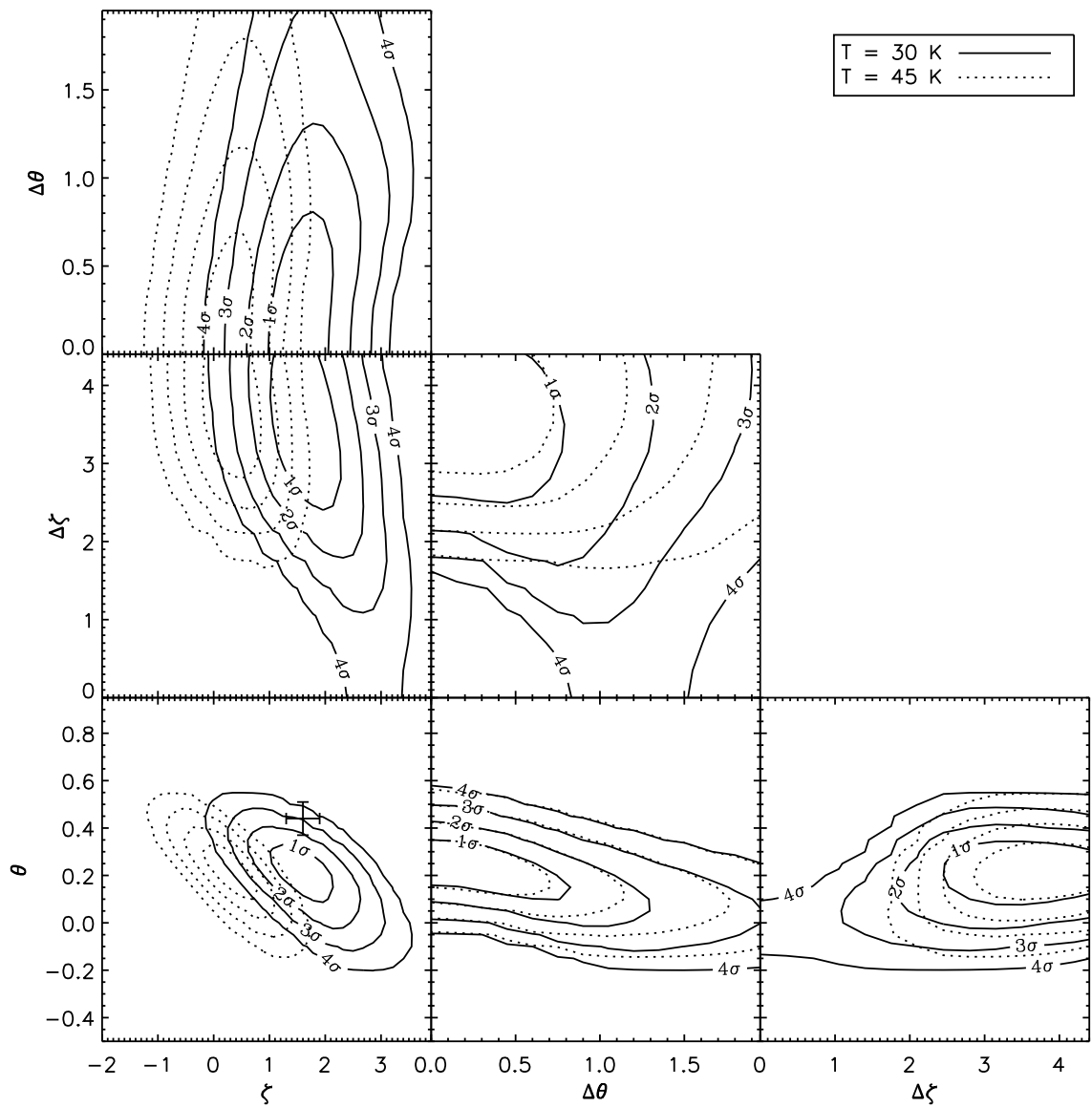}
\caption{Likelihood contours for pairs of the parameters $\theta$ and $\zeta$ (from equation \ref{eq:lz}) and their intrinsic dispersions $\Delta{\theta}$ and $\Delta{\zeta}$, marginalised over the parameters not displayed in each panel. Contours labelled 1-$\sigma$, 2-$\sigma$, 3-$\sigma$, and 4-$\sigma$ enclose 68.3, 95.4, 99.73, and 99.994 per cent of the total likelihood.  The point with error bars in the lower left-hand panel shows the average values of $\theta$ and $\zeta$, and their 1-$\sigma$ uncertainties, determined by \citetalias{SH09} via stacking in the mid-infrared.  Solid contours show a fit to objects assuming $T=30$~K; dotted contours show a fit to the objects assuming $T=45$~K. }\label{fig:likecontours}
\end{figure*}

Figure \ref{fig:likecontours} shows the results of the likelihood evaluations, 
marginalised over all parameters except the two shown in each panel.  This is the result of averaging the likelihoods obtained in 17 random realisations of the background, as described in Section \ref{sec:data}.  Contours
labelled 1-$\sigma$, 2-$\sigma$, 3-$\sigma$, and 4-$\sigma$ enclose 68.3, 95.4, 99.73, and 99.994 per cent of the total likelihood.  We show the results of fits assuming both $T=30$~K and $T=45$~K to give an indication of the way in which our uncertainty in $T$ should contribute to the uncertainty in the derived parameters.  

This figure shows that, assuming the dust in quasars has a temperature $T=30$~K,
 there is good evidence for correlations 
of $L_\mathrm{IR}$ with both $L_{\mathrm{QSO}}$ and $z$, since the 
maximum likelihood values of both $\theta$ and $\zeta$ are significantly
non-zero, and that while higher temperatures can mimic the effect of the correlation with $z$, the correlation with $L_{\mathrm{QSO}}$ is robust to simple changes in the overall dust temperature.  For the $T=30$~K case, the best-fitting values of each 
parameter, with their 68.3 per cent 
confidence limits  (marginalising over all other parameters in each case) are:

\begin{eqnarray}
\theta         & = &  0.22 \pm\, 0.08 \\
\zeta          & = & 1.6 \pm\,  0.4  \\
\Delta{\theta} & = & 0.08 \pm\,  _{0.08}^{0.48}  \\
\Delta{\zeta}  & = & 3.4 \pm\,  _{0.7}^{0.9}  
\end{eqnarray}

Although, as can be seen in the lower left panel of Figure \ref{fig:likecontours}, there is some degeneracy between $\theta$ and $\zeta$, our data exclude $\theta=0$ at approximately the 2.5-$\sigma$ level.   The (flat) prior on $\theta$ allowed it to take values between -1.5 and 1 (i.e. a much larger range than we show in Figure \ref{fig:likecontours}), so we are confident that this is not simply an artefact of reaching the edge of the prior range.

The other panels of Figure \ref{fig:likecontours} show that there is not a
strong degeneracy between any other parameters, so it is not possible
to reproduce the effect of a correlation of 
$L_\mathrm{IR}$ with either $L_{\mathrm{QSO}}$ or $z$ by simply increasing the 
intrinsic dispersion in the other parameter; indeed these panels suggest 
that approximately the same result would be obtained if dispersion 
in the power law indices was ignored altogether.

While the figure shows that the detected evolution of $L_\mathrm{IR}$ with redshift is very strongly dependent on the assumed value of $T$ (although a temperature as high as 45~K for the whole sample seems unlikely), the correlation with $L_{\mathrm{QSO}}$ seems to be robust to this uncertainty.  We find a
slightly weaker index, $\theta$, 
 for the correlation with $L_{\mathrm{QSO}}$ than found by
\citetalias{SH09}, whose average result ($\theta = 0.44 \pm 0.07$) is shown in figure 
\ref{fig:likecontours} as 
the point with errorbars.  This may in part be due to the relatively 
large redshift bins (e.g. $0.5<z<1$, $1<z<2$) which were used in SH09, 
since, because of the strong degeneracy between $\theta$ and $\zeta$, these could allow trends with redshift to affect objects within a single bin.  However, the
disagreement is not very significant.  

\begin{figure*}
\includegraphics[angle=0,width=0.95\textwidth,clip]{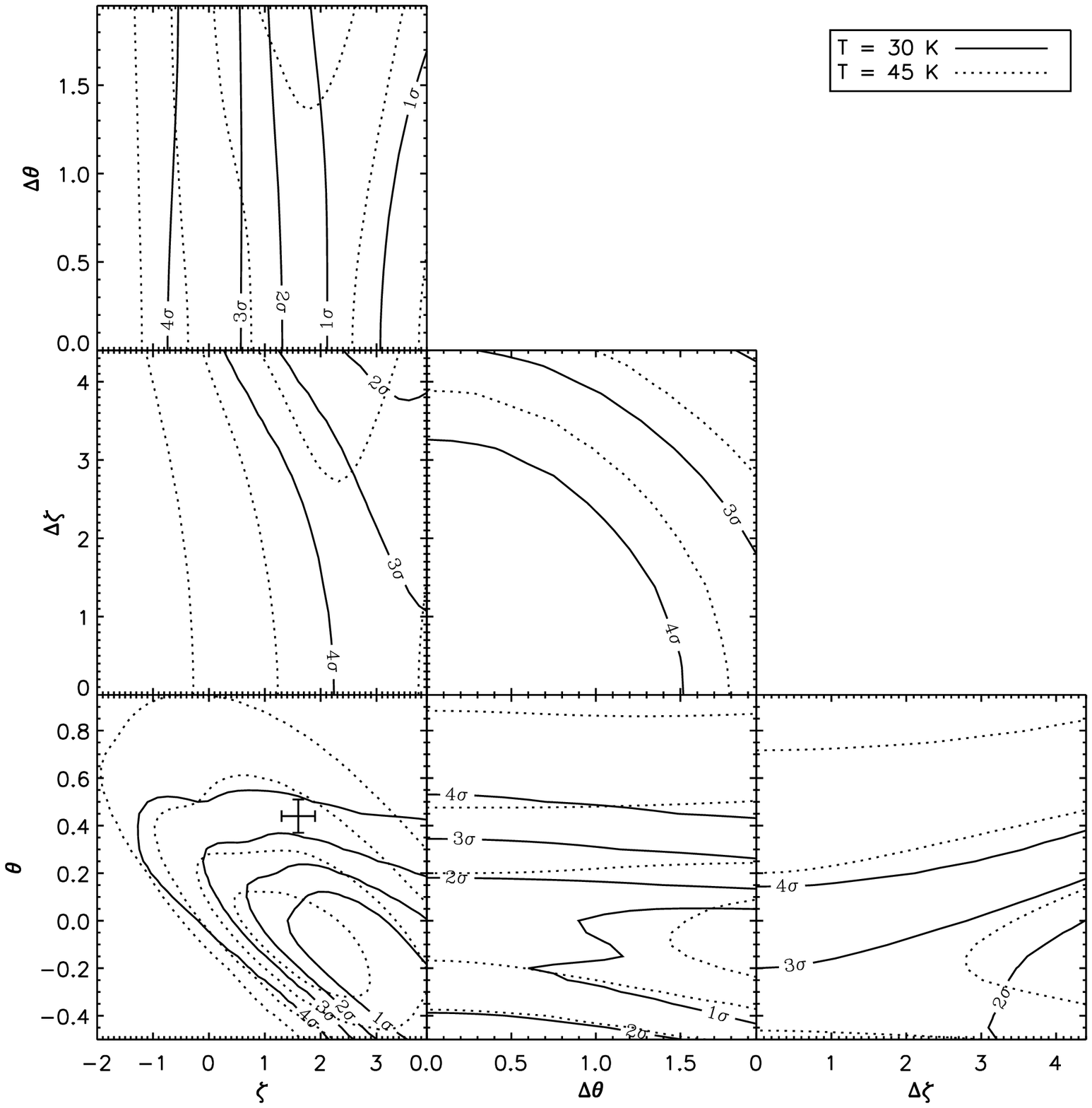}
\caption{As Figure \ref{fig:likecontours} but with infrared fluxes drawn from random positions in the maps rather than from the quasar positions, as a check that the noise properties in the maps are not responsible for the correlation we detect between $L_\mathrm{IR}$ and $L_{\mathrm{QSO}}$.}\label{fig:random}
\end{figure*}

In contrast, performing the same maximum-likelihood fit to data with far-infrared fluxes drawn from random positions in the maps (with the same position used at each wavelength, and the dust-fitting performed assuming a temperature of 30 K) we obtain the results shown in Figure \ref{fig:random}.  The important thing to note is that the best-fitting value of $\theta$ is zero for the random flux case, indicating that the result we have obtained is not simply due to the noise properties of the far-infrared maps, but rather is due to far-infrared flux genuinely associated with the quasars.  (Note that the non-zero value for $\zeta$ derived from the fit to random data is not unexpected, since the real quasar luminosities and redshifts were still used in this case, and these are themselves correlated.)

Using deeper far-infrared data from \emph{Herschel}, but with a narrower range in optical luminosity (because only SDSS, and not 2SLAQ, quasars are available in the field), \citet{hatz10} find a very similar result, $\theta=0.35$, for objects at redshifts $z>2$.  \citet{serjeant2010} use a compilation of H-ATLAS and other data to find a weakly declining correlation between $L_\mathrm{IR}$ and quasar luminosity, parametrised as $\theta = (0.5875 \pm 0.045) - (0.09275 \pm 0.012)\,z$ for redshifts $0<z<4$, which is a slightly higher value of $\theta$ than we find, but broadly consistent within the errors and at the redshifts we are sensitive to.  Our value of $\theta$ is also consistent with the work by \citet{silverman2009}, who find $\theta=0.28 \pm 0.22$ for the slope of the power law relating [O{\sc ii}] luminosity (tracing star formation after correcting for AGN contamination) to X-ray luminosity (tracing the AGN). {  Silverman et al.~study a sample of AGN over a narrower range of redshifts ($0.5<z<1$) and with somewhat lower luminosities than most of our sample ($10^{44.3}$ erg~s$^{-1} \ltsim L_{\mathrm{bol}} \ltsim 10^{45.8}$ erg~s$^{-1}$, assuming a bolometric correction $L_{\mathrm{bol}} / L_{\mathrm{2-10 keV}} \sim 30$); the consistency of their result with ours suggests that the relationship between $L_\mathrm{IR}$ and $L_{\mathrm{QSO}}$ may not vary with $L{\mathrm{QSO}}$ as suggested by \citet{lutz2010} and \citet{shao2010}.  Their steeper power law, with a slope of 0.8, for luminous  quasars is not directly comparable with our result as it does not separate out the independent effect of redshift on $L_\mathrm{IR}$.  However, we would also note that the quasar part of this relationship is based on samples presented by \citet[][and earlier works]{netzer2009}, which ignore AGN without detections in the far-infrared or mm, so that the observed correlation may be partly due to far-infrared flux limits.}

\begin{figure}
\includegraphics[angle=0,width=0.48\textwidth,clip]{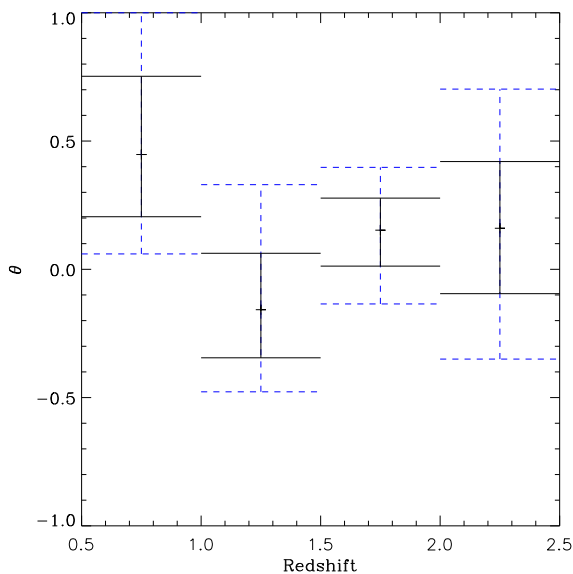}
\caption{Best-fitting values of $\theta$ based on maximum-likelihood fits using data in the redshift ranges denoted by the widths of the errorbars.  Black, solid errorbars are 68.3 per cent confidence limits; blue, dashed errorbars are 95.4 per cent confidence limits.  }\label{fig:lcoeff}
\end{figure}

To investigate whether our fit could be affected by a redshift-dependence in $\theta$, we perform the maximum likelihood fitting again, using objects in restricted redshift ranges where we have enough objects to make a robust measurement, and marginalise over all parameters except $\theta$.  The results of this are shown in Figure \ref{fig:lcoeff}; the derived values of $\theta$ are not significantly different from each other for redshifts $0.5<z<2.5$, although the small number of objects in the redshift bins means we cannot make a strong statement for or against evolution of $\theta$ as a function of redshift.  However, the $0.5<z<1$ range shows a detection of non-zero $\theta$ at better than the 2-$\sigma$ level, which is complementary to the result of \citet{hatz10}.  While we do not have enough data to detect any significant decline in $\theta$ as a function of redshift, the datapoints in Figure \ref{fig:lcoeff} are consistent with the relation found by \citet{serjeant2010}, and with our smaller H-ATLAS-only dataset we cannot rule out the existence of such a weak redshift dependence.  

Since the far-infrared luminosity $L_\mathrm{IR}$ has been shown to be due to 
star formation \citep{hatz10}, the physical interpretation of these results
is that star-formation rate is higher both as redshift increases and as 
quasar luminosity increases.  As was also suggested by \citetalias{SH09}, 
and as has been incorporated into the ``quasar-mode'' feedback recipes of 
semi-analytic models \citep[e.g.][]{croton2006}, 
{ one of the simplest ways} to interpret 
this is if both black hole
accretion and star formation depend on a common supply of cold gas, as 
might be supplied during a galaxy--galaxy merger or cold-accretion event 
\citep{dekel2009,dijkstra2009,smith2007,smith2008}.  { The implication of the 
non-unity-slope is that the dependence of either star-formation or accretion 
luminosity (or both) on the gas supply cannot simply be linear with gas mass, 
as one might na{\"\i}vely suppose, but is instead more complex}.

For both star formation and black-hole accretion, such gas must  
be sufficiently cool that pressure support 
does not prevent it
from being accreted by the AGN or from collapsing 
under its own gravity to form stars,
so an alternative way to make this connection is via a mechanism (such as AGN 
or supernova feedback) which is capable of heating gas at all scales within a galaxy.

\section{Conclusions}\label{sec:conc}
Using far-infrared measurements (at 250, 350 and 500\um{}) of SDSS and 2SLAQ quasars in the H-ATLAS science demonstration field, we find evidence for correlations between the infrared luminosities of quasar hosts with both redshift and quasar luminosity.  While the best-fit power-law index for the correlation with redshift can vary significantly depending on the assumed temperature of the dust in the quasar hosts, the correlation with quasar luminosity is robust to this effect, and is non-zero at the 2.5-$\sigma$ level when we combine all objects in a maximum-likelihood fit.  

We find a power-law index $\theta=0.22\pm 0.08$ for the relationship between $L_{\mathrm{IR}}$ and $L_{\mathrm{qso}}$.  Our data show that $\theta$ is non-zero at the 2-$\sigma$ level in the $0.5<z<1$ redshift range alone, but we have too few objects to find evidence for evolution in $\theta$ over the range $0.5<z<2.5$.   

We find evidence for a large intrinsic dispersion in the redshift dependence, but no evidence for intrinsic dispersion in the correlation between $L_{\rm QSO}$ and $L_{\rm IR}$, suggesting that the latter is due to a direct physical connection between star formation and black hole accretion.  { One possible interpretation of this is that} both the quasar activity and star formation are dependent on the same reservoir of cold gas, and are thus both affected by influx of gas during mergers or cold accretion, or heating of gas via feedback processes.

With the full 550 square degrees of the H-ATLAS survey, we will be able to substantially increase the sample of quasars and improve these constraints.  The larger area will allow us to obtain representative samples of the objects with the lowest density on the sky, i.e. quasars with high luminosities at low redshift, which will increase the sensitivity of our test at low redshifts and permit us to remove the assumption of power-law correlations and test more general models.  More importantly, a larger sample of objects at all redshifts will make it possible to measure the temperatures of dust in quasar hosts as a function of $L_{\mathrm{qso}}$ and redshift by stacking the far-infrared fluxes; this in turn will remove the largest source of uncertainty in the relationship between $L_{\mathrm{IR}}$ and $z$.

\section*{Acknowledgements}
MJJ acknowledges support from an RCUK fellowship.  MJH thanks the Royal Society for a Research Fellowship.

The \emph{Herschel}-ATLAS is a project with \emph{Herschel}, which is an ESA space observatory with science instruments provided by European-led Principal Investigator consortia and with important participation from NASA. The H-ATLAS website is http://www.h-atlas.org/.

%The Herschel spacecraft was designed, built, tested, and launched under a contract to ESA managed by the Herschel/Planck Project team by an industrial consortium under the overall responsibility of the prime contractor Thales Alenia Space (Cannes), and including Astrium (Friedrichshafen) responsible for the payload module and for system testing at spacecraft level, Thales Alenia Space (Turin) responsible for the service module, and Astrium (Toulouse) responsible for the telescope, with in excess of a hundred subcontractors.

Funding for the SDSS and SDSS-II has been provided by the Alfred P. Sloan Foundation, the Participating Institutions, the National Science Foundation, the U.S. Department of Energy, the National Aeronautics and Space Administration, the Japanese Monbukagakusho, the Max Planck Society, and the Higher Education Funding Council for England. The SDSS Web Site is http://www.sdss.org/.

The SDSS is managed by the Astrophysical Research Consortium for the Participating Institutions. The Participating Institutions are the American Museum of Natural History, Astrophysical Institute Potsdam, University of Basel, University of Cambridge, Case Western Reserve University, University of Chicago, Drexel University, Fermilab, the Institute for Advanced Study, the Japan Participation Group, Johns Hopkins University, the Joint Institute for Nuclear Astrophysics, the Kavli Institute for Particle Astrophysics and Cosmology, the Korean Scientist Group, the Chinese Academy of Sciences (LAMOST), Los Alamos National Laboratory, the Max-Planck-Institute for Astronomy (MPIA), the Max-Planck-Institute for Astrophysics (MPA), New Mexico State University, Ohio State University, University of Pittsburgh, University of Portsmouth, Princeton University, the United States Naval Observatory, and the University of Washington.

\bibliographystyle{mn}
\bibliography{qsos}
\bsp

\label{lastpage}

\end{document}